# Linear Time Split Decomposition Revisited

Pierre Charbit* Fabien de Montgolfier* Mathieu Raffinot

LIAFA, Univ. Paris Diderot - CNRS, ANR Project Graal
{charbit,fm,raffinot}@liafa.jussieu.fr

**Abstract**

Given a family $\mathcal{F}$ of subsets of a ground set $V$, its orthogonal is defined to be the family of subsets that do not overlap any element of $\mathcal{F}$. Using this tool we revisit the problem of designing a simple linear time algorithm for undirected graph split (also known as 1-join) decomposition.

## 1 Introduction

Given a family of subsets of a ground set $V$, its orthogonal is defined to be the family of subsets that do not overlap any element of $\mathcal{F}$. The computation of the orthogonal of a general family $\mathcal{F}$ has been made linear by R. McConnell in [15] in which it is the core of a linear time algorithm to test the consecutive ones property of $\mathcal{F}$. The purpose of this article is to explain how the orthogonal tool can be successfully applied to design a simple linear time split (or 1-join) decomposition of undirected graphs.

Let us rapidly survey the notion of graph decomposition in general and the particular decomposition we are interested in. The main idea of graph decomposition is to represent a graph by a simpler structure (usually a tree) that is built and labelled in such a way that some properties of the graph we are interested in are embedded in the structure. Solving a problem on the graph might then be done by just manipulating its decomposition, using dynamic programming for instance, which usually leads to simple and fast algorithms. Many graph decompositions exist and some are well known, like for instance the decomposition by clique separators [25] or the modular decomposition [26, 17, 20].

The split decomposition, also known as 1-join decomposition, is one of those famous decomposition that has a large range of applications, from NP-hard optimization [23, 22] to the recognition of certain classes of graphs such distance hereditary graphs [10, 11], circle graphs [24] and parity graphs [4, 8]. A survey on applications of the split decomposition in graph theory can be found in [23]. This decomposition has been introduced by Cunningham in [6] who also presented a first worst case $O(n^3)$-time algorithm. This complexity has been improved to $O(nm)$ in [9] and to $O(n^2)$ in [14] ($n$ being the number of vertices and $m$ the number of edges of the graph).

Two papers have been written by E. Dahlhaus for solving the problem in linear time: an extended abstract in 1994 [7] followed several years later only (in 2000) by an article in Journal of Algorithms [8]. However, while these two last manuscripts substantially differ, they are both very difficult to read, and the algorithm presented is such involved that its proof and linear-time complexity can hardly be checked.

*INRIA project-team GANG

The notion of orthogonal computed using McConnell linear time algorithm allows us to go deeper in the understanding of the structure of the splits of a graph which at its turn is the key to obtain a more comprehensive and well founded linear time split decomposition algorithm. The paper is organized as follows. In the next section we present the notion of orthogonal that is closely linked to partitives families. Section 3 is devoted to theoretical aspects of split decomposition and in Section 4 we prove our new algorithm based on orthogonals. Its complexity is stated in the last section.
In the remaining of the paper, $V$ denotes a finite set, $2^V$ denotes the set of all subsets of $V$. For a family $\mathcal{F} \subset 2^V$, we define its norm $\|\mathcal{F}\| = |\mathcal{F}| + \sum_{X \in \mathcal{F}} |X|$ . We also always assume that our families satisfy two properties which do not imply any loss of generality in our context: (a) $V \in \mathcal{F}$, and (b) $\forall x \in V,\ \{x\} \in \mathcal{F}$. The following two notions are of main importance in this paper.

**Definition 1** *Two subsets of* $V$ overlap *if their intersection is non empty but none is included in the other. If two subsets* $X$ *and* $Y$ *of* $V$ *do not overlap, we say they are* orthogonal *which is denoted* $X \perp Y$.

## 2 Partitive Families and Orthogonals

In this first section, we recall and detail a problem that is related to the following very general question: if $V$ is a finite set, which families of subsets of $V$ have a compact representation and for which are we able to compute it?

To illustrate the previous question, let us start with a very simple example. Assume our family $\mathcal{F}$ satisfies the following:

$$\forall X, Y \in \mathcal{F},\ X \perp Y$$

This type of family is called *laminar*. No two elements of $\mathcal{F}$ overlap, and thus it is straightforward to notice that such a family can be represented by a rooted tree, such that

- the leaves of this tree are in bijection with elements of $V$;
- the nodes of this tree are in bijection with elements of $\mathcal{F}$ the following way: each node of the tree represents the subset of $V$ consisting of all elements corresponding to leaves of the subtree rooted by this node.

Figure 1 illustrates this simple example. Partitive families are a more evolved example.

**Definition 2** *A family* $\mathcal{F}$ *of subsets of* $V$ *is partitive if*

- $V$ *and all singletons belong to* $\mathcal{F}$
- *for all* $X, Y \in \mathcal{F}$ *such that* $X$ *overlaps* $Y$, $X \cup Y$, $X \cap Y$, *and* $(X \setminus Y) \cup (Y \setminus X)$ *are also in* $\mathcal{F}$.

Notice that laminar families are a special kind of partitive families, and the tree representation of partitive families that follows generalizes that of laminar.

A partitive tree is a rooted tree $T$ whose node are labelled *Prime* or *Complete*, and whose leaves are labelled in bijection with the elements of $V$. We associate to such a tree the family of subsets of $V$, that are of two kinds:

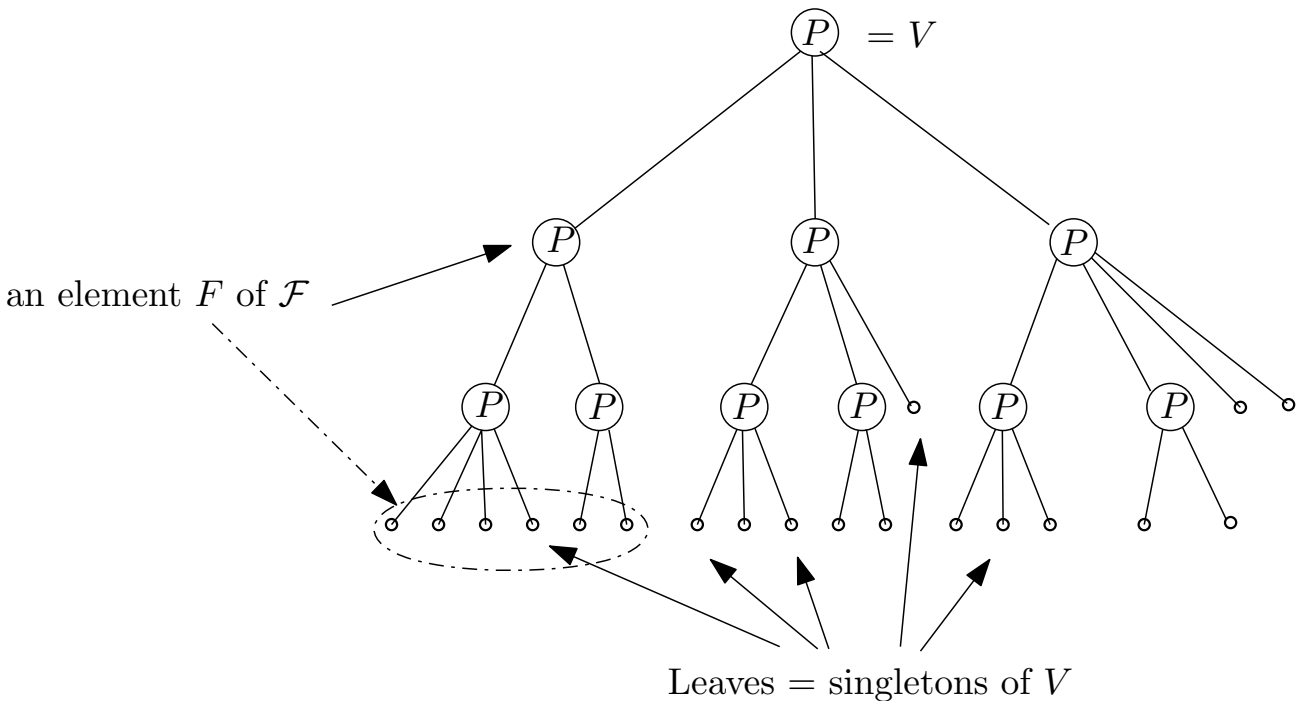

*Figure 1: Tree representation of a laminar family.*

- for every Prime node of the tree, the subset of $V$ consisting of all elements corresponding to leaves of the tree that are descendants of this node,
- for every Complete node, and for every possible union of its children, our family contains the union of the subsets of $V$ represented by these children.

Figure 2 shows a partitive tree. The following theorem states that every partitive family can be represented this way.

**Theorem 1** *[3, 19, 13] Any partitive family can be represented by a partitive tree.*

Note that there is an ambiguity on nodes with two sons, they may me labelled Prime or Complete, in this paper we always choose to label these Complete.

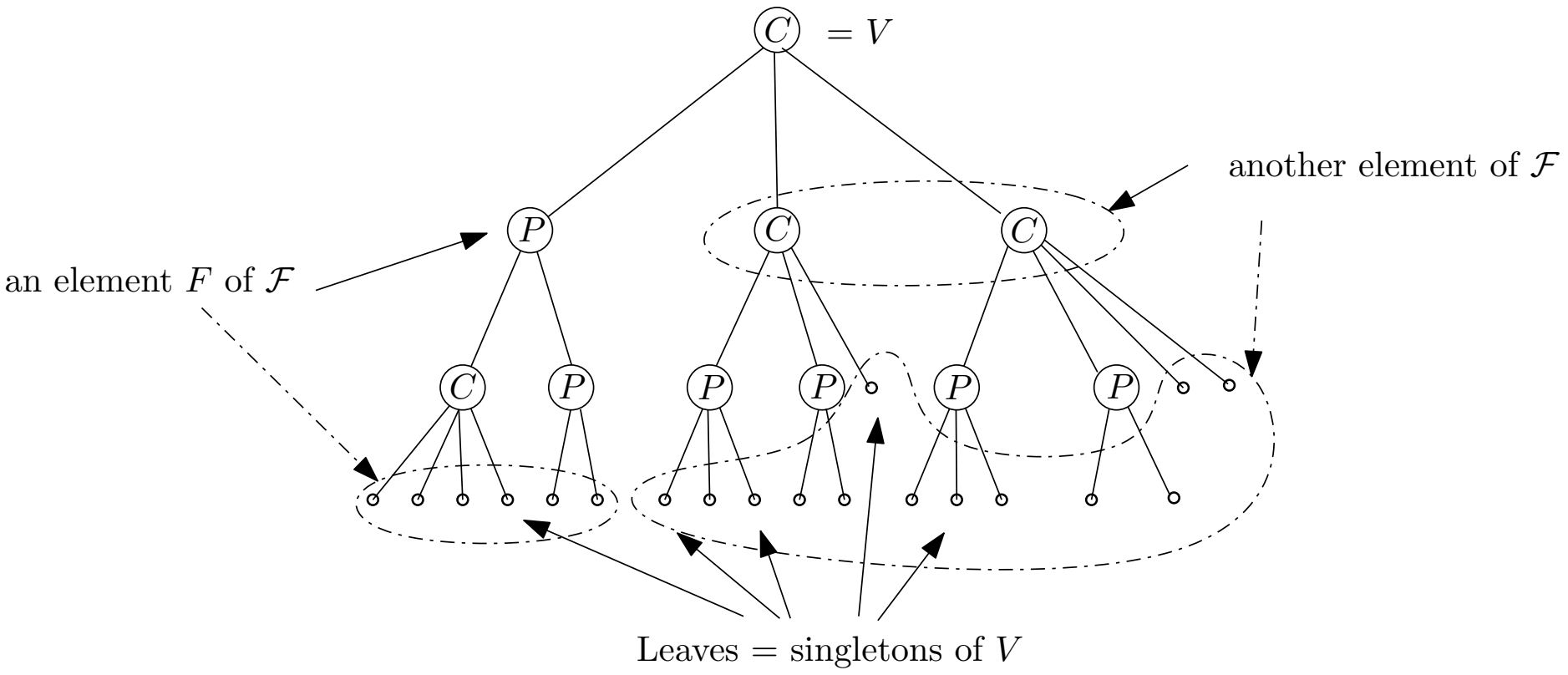

*Figure 2: Partitive family*

We present now another way of seeing partitive families. It is related to the central notion of this section, that is defined below.

**Definition 3** *Let $\mathcal{F} \subset 2^V$ be a family of subsets of $V$. Its* orthogonal*, denoted by $\mathcal{F}^\perp$, is defined by*

$$\mathcal{F}^\perp = \{X \subseteq V,\ \forall Y \in \mathcal{F},\ X \perp Y\}.$$

The following results are easy.

**Proposition 1** $(\mathcal{F} \cup \mathcal{F}')^\perp = \mathcal{F}^\perp \cap \mathcal{F}'^\perp$

**Proposition 2** *$\mathcal{F}^\perp$ is a partitive family.*

**Proposition 3** *If $\mathcal{F}$ is partitive, then the tree representation of $\mathcal{F}^\perp$ is obtained from the one of $\mathcal{F}$ by switching* Prime *and* Complete *nodes.*

**Corollary 1** *If $\mathcal{F}$ is partitive, then $\mathcal{F}^{\perp\perp} = \mathcal{F}$. Therefore, every partitive family $\mathcal{F}$ is the orthogonal of some family $\mathcal{F}'$*

After these definitions, we can expose the main result in this section. Given a general family $\mathcal{F}$, the following theorem of McConnell ([15]) states that it is possible to compute the tree representation of its orthogonal in an efficient way.

**Theorem 2** *[15] Given a family of subsets $\mathcal{F}$, it is possible in $O(\|\mathcal{F}\|)$ time to compute the partitive tree representation of $\mathcal{F}^\perp$.*

It should be noticed that the linear time algorithm in [15] for computing the orthogonal of a general family $\mathcal{F}$ is mainly based on an algorithm of Dahlhaus for computing overlap classes presented in [8], that has been recently revisited, simplified, extended and implemented in [2, 21]. The main computational insight is that although the overlap graph of $\mathcal{F}$ can be of quadratic size, overlap components can be computed in $O(\|\mathcal{F}\|)$ time.

# 3 Split Decomposition - Theory

In this section, we show how Theorem 2 can be used as the main ingredient to compute the split decomposition of undirected graphs. In the rest of the paper, $G = (V, E)$ denotes a simple connected graph. For $X \subset V$, we denote by $N(X)$ its set of neighbors, that is the set of vertices $y \notin X$ such that there exists $xy \in E$ with $x \in X$.

## 3.1 Introduction

We now recall some definitions and previous results on splits, and we define precisely the structure we are aiming for. Some proofs are omitted, we refer the reader to the pioneering work of Cunningham (see [6] for more details).

**Definition 4** *A split of $G = (V, E)$ is a bipartition of $V = X_1 \cup X_2$ such that the edges between $X_1$ and $X_2$ induce a complete bipartite graph.*
*In other words, there exists a partition of $V$ into 4 subsets $V_1, V_2, V_3, V_4$, such that $X_1 = V_1 \cup V_2$ and $X_2 = V_3 \cup V_4$, and such that $G$ contains all edges between $V_2$ and $V_3$, and no other edges between $X_1$ and $X_2$.*
*We denote splits either by bipartitions $(X_1, X_2)$ or by quadripartitions $(V_1, V_2, V_3, V_4)$ depending on needs. Both are equivalent since there is an unique quadripartition for each bipartition.*
*A split is said to be* non trivial *if both sides have more than two vertices.*

Figure 3 illustrates the notion of split. A special case of split is the notion of module.

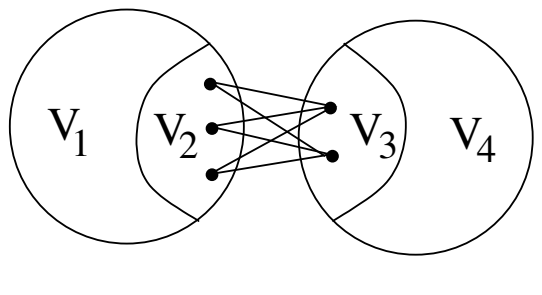

*Figure 3: Structure of a split.*

**Definition 5** *A subset $M$ of $V$ is called a module if, using the notations of the previous definition, there exists a split such that $V_1 = \emptyset$ and $M = V_2$. A module is* strong *if it overlaps no other module.*

Modules, also called homogeneous sets, appear in various contexts, as for example perfect graphs, claw free graphs or design of efficient algorithms (see for instance [20, 1]). Their structure is well studied, and the representation of all modules is a tree called modular decomposition. Given a graph $G$ there exist linear $O(|V|+|E|)$-time algorithms to compute this decomposition [5, 18].

A graph may contain an exponential number of splits. For instance in a complete graph every bipartition is a split (in fact a module). However all splits may be represented in a compact way, this is where the notion of strong split appears.

**Definition 6** *Two splits $(V_1, V_2, V_3, V_4)$ and $(V_1', V_2', V_3', V_4')$* cross *if $V_1 \cup V_2$ overlaps both $V_1' \cup V_2'$ and $V_3' \cup V_4'$. A split if* strong *if it crosses no other split.*

What is fundamental is that the strong splits have a simple, partitive-like, structure.

**Theorem 3** *Fix $r \in V(G)$.*

$$\{X \subset V(G),\ (X, V \setminus X) \text{ is a strong split of } G \text{ and } r \notin X\} \text{ is a partitive family of } V(G)$$

Therefore, adding just an edge with leaf $r$ at the root of the partitive tree representing this family yields an unrooted tree which almost represents all splits of $G$. Indeed, its leaves are in bijection with $V$ and its set of edges of $T$ are in bijection with the strong splits of $T$ in the following way : to each edge $e$ of $T$ is associated the bipartition of $V$ given by the labels of the leaves of the two connected components of $T - e$. We now label the nodes of this tree to represent all splits.

**Definition 7** *Suppose $V(G)$ admits a partition $(V_1, V_2, \ldots, V_k)$ such that each $(V_i, V(G) \setminus V_i)$ defines a split of $G$. Construct a graph $H$ with $k$ vertices $(v_1, v_2, \ldots, v_k)$ and with an edge $v_i v_j$ if and only if $G$ contains an edge between $V_i$ and $V_j$.*

*The graph $H$ is called the* quotient graph *with respect to this partition into splits.*

An important result is the following.

**Proposition 4** *Suppose $V(G)$ admits a partition $(V_1, V_2, \ldots, V_k)$ such that each $(V_i, V(G) \setminus V_i)$ defines a split of $G$. Let $H$ be the quotient graph associated to this partition.*

*If $A = \{v_i,\ i \in I\}$ defines a non trivial (resp. strong) split $(A, V(H) \setminus A)$ of $H$ for some $I \subset \{1, \ldots, k\}$, then $(\cup_{i \in I} V_i, \cup_{i \notin I} V_i)$ is a non trivial (resp. strong) split of $G$.*

What it implies is that if we dispose of a tree representing all strong splits of the graph, there are only few possible cases for the nodes. Using the notations of the previous proposition, either the quotient graph $H$ is a prime graph (it contains no non trivial splits), and in which case no union of the $V_i$ defines a split of the graph, or it contains some non trivial splits but no strong one. In the latter case it is not difficult to prove that the graph $H$ is either a clique $K_n$, or a star $K_{1,n}$. In both cases all possible unions of the $V_i$ define a split with respect to their complement. Figure 4 shows an example of these three cases.

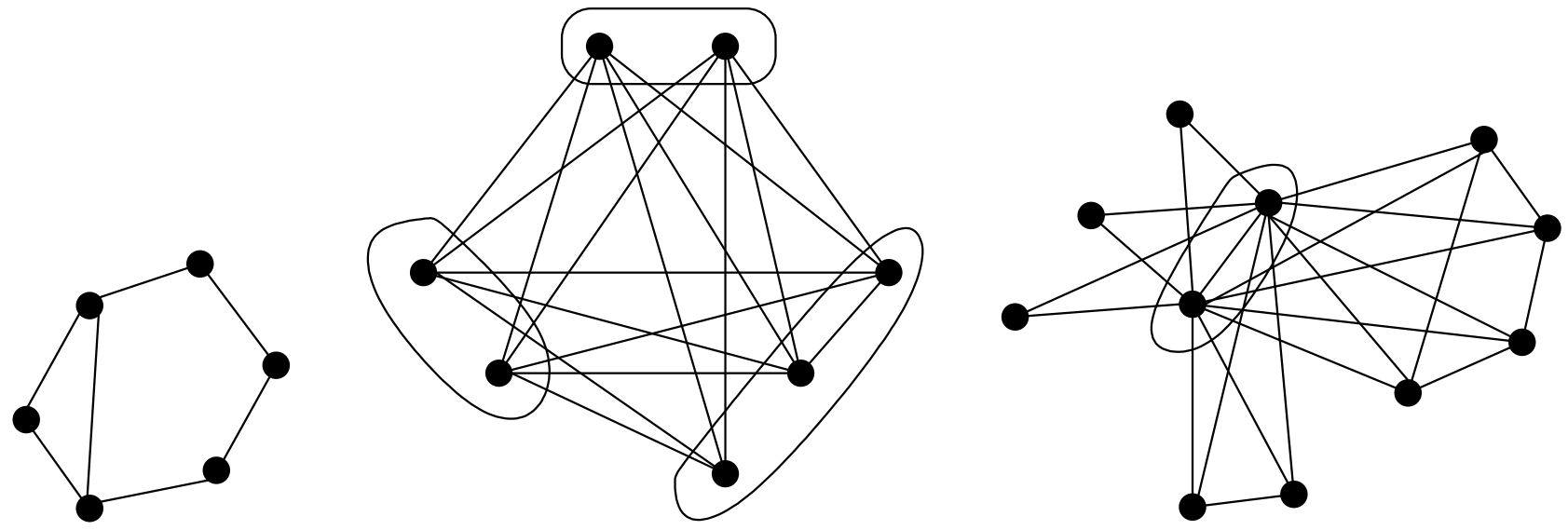

(a) A prime subgraph (b) A clique subgraph (c) A star subgraph

*Figure 4: Examples of the three distinct type of nodes of Cunningham's tree. The center of the star is encircled.*

The whole tree, with its nodes labelled Prime, Star or Clique (corresponding to the three possible cases described in the previous paragraph) and the orientation associated to each Star node to points its center, forms the Cunningham's tree decomposition and is the structure we are building in the rest of the paper. An example of such a tree is given in Figure 5. Our approach is first to fix a vertex $r$ as a root in our graph and then use Theorem 3 to find all parts of strong splits that do not contain $r$. The previous discussion implies the following result of Cunningham.

**Proposition 5** *[6] Let $(X,Y)$ and $(X',Y')$ be two crossing splits, there exists either a Star $(V_1, V_2, \ldots, V_k)$ or a Clique $(V_1, V_2, \ldots, V_k)$ and $\emptyset \subsetneq I, I' \subsetneq \{1, \ldots, k\}$ such that $X = \cup_{i \in I} V_i$ and $X' = \cup_{i \in I'} V_i$.*

## 3.2 Split Borders

Let $r$ be a vertex of $G$. For each split $(V_1, V_2, V_3, V_4)$ (using notations of Definition 4), $r$ lies either in $V_1 \cup V_2$ or in $V_3 \cup V_4$. Without loss of generality, we consider that for all splits we have $r \in (V_1 \cup V_2)$. The root vertex $r$ then allows us to "orient" every split.

**Notations:** The set $V_3 \cup V_4$ is called the *split bottom* and the set $V_3$ is called the *border* of the split $(V_1, V_2, V_3, V_4)$. Notice that two different splits bottoms may share the same border. We define the *distance* of a split bottom (resp. border) $S$ as its distance from the root, that is $\min_{x \in S} d(r, x)$. We denote $G[h]$ as the subgraph induced by vertices at distance $h$, and $G[\leq h]$ as the subgraph induced by vertices at distance at least $h$, and similarly $G[< h]$ or $G[> h]$ in the obvious way. For $X \subset G[> h]$ we denote $N_h(X)$ the set $N(X) \cap G[h]$. Moreover, the letter $H$ always denotes the set of vertices of $G[h]$. Note also that all orthogonal notations here refer to the orthogonal with respect to the ground set $H$.

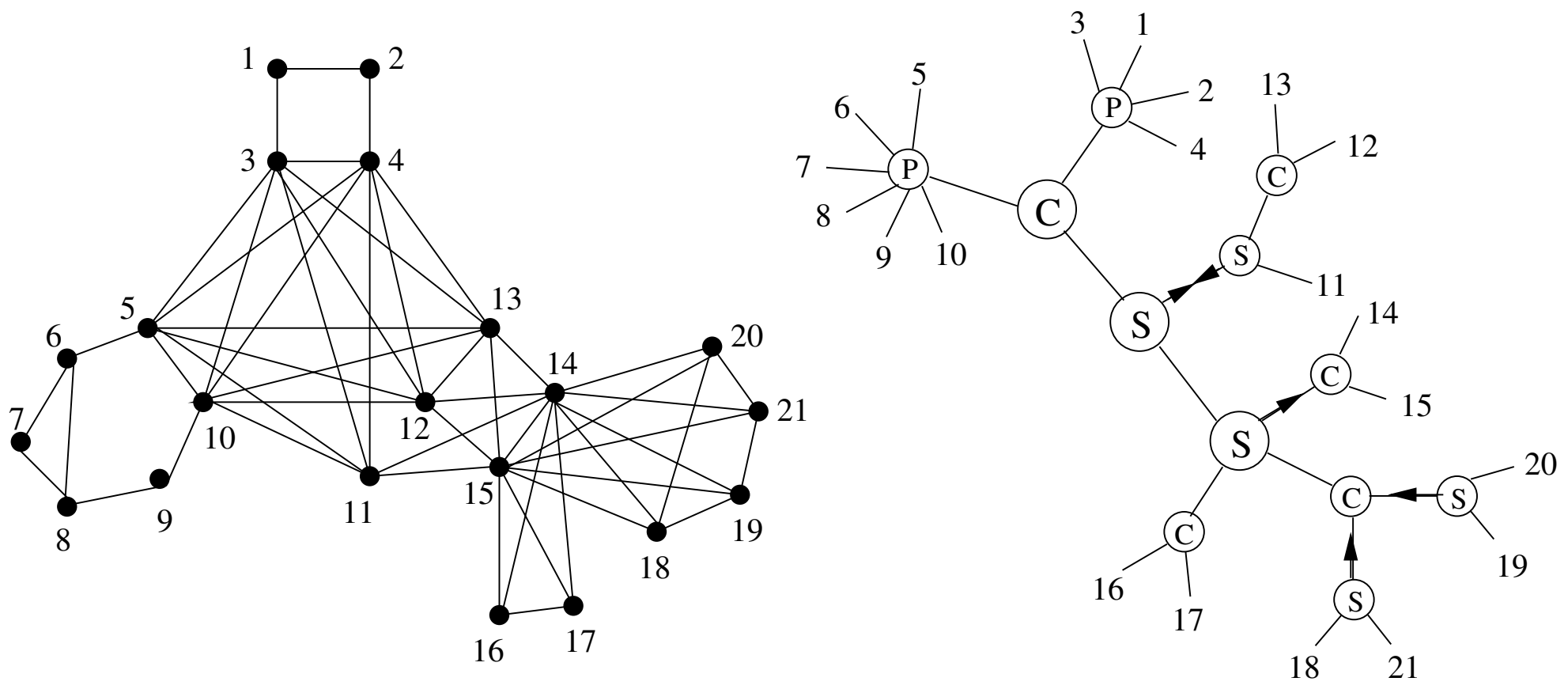

*Figure 5: An example of graph and its corresponding* split tree. *Nodes labelled C, S and P are respectively clique, star and prime. An orientation is associated to each star node to point its center. Note that nodes with 3 incident edges could have been labelled* prime

**Lemma 1** *All vertices of a given border are at the same distance from the root $r$.*

This justifies the approach of our algorithm: we first compute (using a bread first search for example) the distance layers of our graph, and then we process one layer after the other in a bottom-up approach from the furthest layer to the first one. At each step, we need to identify the set of borders at distance $h$ from the root $r$.

Let us denote by $\mathcal{B}_h$ the set of all borders of split at distance $h$ from the root $r$. Let $C_1, \ldots, C_k$ be the connected components of $G[> h]$. We define two families of subsets of $H$:

- $\mathcal{M} = \{$modules of $G[\leq h]$ that are subsets of $H\}$,
- $\mathcal{V} = \bigcup_{i=1}^{k} \mathcal{V}_i$ , where

$$\mathcal{V}_i = \{N(C_i) \cap H\} \cup \{N(x) \cap H, x \in C_i\} \cup \{(N(C_i) \setminus N(x)) \cap H, x \in C_i\}$$

**Theorem 4**

$$\mathcal{B}_h = \mathcal{M} \cap \mathcal{V}^{\perp}$$

**Proof** We use the notations $(V_1, V_2, V_3, V_4)$ of Definition 4 to denote the different parts of a split. Let $B = V_3$ be an element of $\mathcal{B}_h$. Since $V_4$ is included in $G[> h]$, $B$ is in fact a module of $G[\leq h]$. Thus $B$ has to be an element of $\mathcal{M}$.
Consider now a connected component $C_i$. Either it is included in $V_4$ in which case all elements of $\mathcal{V}_i$ are clearly subsets of $B$, or it is included in $V_1 \cup V_2$. In the latter case, vertices in $V_1$ have no neighbor in $B$ and vertices in $V_2$ see all elements of $B$. This implies that $B \in \mathcal{V}^{\perp}$.

Conversely, assume $B \in \mathcal{M} \cap \mathcal{V}^{\perp}$. Let $V_3 = B$ and let $V_4$ be the union of all $C_i$ of $G[> h]$ such that $(N(C_i) \cap H) \subset B$. Let $V' = V \setminus (V_3 \cup V_4)$. For every $x \in V'$,

- either $x$ belongs to some $C_i$. Since $B \in \mathcal{V}^{\perp}$, $B$ does not overlap $N(x) \cap H$ nor its complement in $N(C_i)$. Thus $x$ sees all vertices of $B$ or no vertex of $B$;
- or $x$ belongs to no $C_i$. Then $x \in G[\leq h]$. Since $B \in \mathcal{M}$, $x$ either sees all vertices of $B$ or no vertex of $B$.

We then define $V_1 = V' \setminus N(B)$ and $V_2 = V' \cap N(B)$. Clearly $(V_1, V_2, V_3, V_4)$ is a split, and thus $B$ is a border. ■

**Theorem 5** *$\mathcal{B}_h \cup \{H\} = (\mathcal{M}^\perp \cup \mathcal{V})^\perp$ and $\mathcal{B}_h \cup \{H\}$ is a partitive family*

**Proof** $\mathcal{M} \cup \{H\}$ is a partitive family. Indeed the union, intersection and symmetric difference of two modules $A$ and $B$ are modules, and if $A$ and $B$ belong to $H$ they do also. However $H$ may fail to be a module, and thus to be a split. To handle this case we apply Proposition 1:

$$\begin{aligned}
\mathcal{B}_h \cup \{H\} &= (\mathcal{M} \cap \mathcal{V}^\perp) \cup \{H\} \quad \text{(by previous theorem)} \\
&= (\mathcal{M} \cup \{H\}) \cap (\mathcal{V}^\perp \cup \{H\}) \\
&= (\mathcal{M} \cup \{H\})^{\perp\perp} \cap (\mathcal{V}^\perp) \\
&= ((\mathcal{M} \cup \{H\})^\perp \cup \mathcal{V})^\perp \quad \text{(by Proposition 1)} \\
&= (\mathcal{M}^\perp \cup \mathcal{V})^\perp
\end{aligned}$$

Proposition 2 gives that $\mathcal{B}_h \cup \{H\}$ is partitive. ■

This theorem is the core of our algorithm to compute Cunningham's split decomposition tree.

### 3.3 Split Bottoms

In the previous section we explicited the structure of the split borders of each layer. We consider now split bottoms, that are related both to split border and to connected components. The following proposition and its corollary below are consequences of the proof of Theorem 4.

**Proposition 6** *Let $B$ be in $\mathcal{B}_h$. If $C$ is a connected component of $G[> h]$, there are only 3 possible cases:*

1. *$N_h(C) \not\subseteq B$. In this case $C$ is not included in any split bottom with border $B$.*
2. *$N_h(C) \subset B$ and there exists $x$ in $C$ such that $\emptyset \subsetneq N_h(\{x\}) \subsetneq B$. In that case $C$ is in every split bottom with border $B$.*
3. *$C$ is a split bottom of distance $h+1$ and $N_h(C) = B$. Then to every split bottom with border $B$ that does not contain $C$, it is possible to add $C$ to get another split bottom with border $B$.*

According to the case, $C$ is said to be of Type 1, Type 2 or Type 3 with respect to $B$.

We call *strong split bottom* the bottom of a strong split. We have:

**Corollary 2** *If $B$ is in $\mathcal{B}_h$, there are only two possible strong split bottoms with border $B$ which are:*

$$B \cup \bigcup_{C \text{ of Type } 2} C \qquad \text{and} \qquad B \cup \bigcup_{C \text{ of Type } 2 \text{ or } 3} C$$

*where $C$ denotes a connected component of $G[> h]$.*

We introduce now a definition and a proposition in order to clarify the link between being a strong border (a border that overlaps no other border) and being a border of a strong split.

**Definition 8** *Let $(V_0, V_1, \ldots, V_k)$, $k \geq 3$ be a partition of $V(G)$ such that each $(V_i, V(G) \setminus V_i)$ defines a split of $G$. If the quotient graph is a star with center $V_0$, and if $r$ does not belong to $V_0$, this partition into splits is called a* bad star *(and a* good star *if $V_0$ contains $r$). If (w.l.o.g.) $r$ belongs to $V_1$, then $\cup_{i \geq 2} V_i$ is clearly a split bottom and its border is called a* bad star border.

**Proposition 7** *If $B$ is in $\mathcal{B}_h \cap \mathcal{B}_h^{\perp}$ (a strong border), then $B$ is the border of a strong split if and only if it is not a bad star border. Furthermore, if $B$ is a bad star border, then no other border strictly contains $B$ and no component of $G[\geq h]$ has an intersection with $G[h]$ that strictly contains $B$.*

**Proof** First suppose that $B$ is a bad star border, at distance $h$ from $r$. The center is $V_0$, $V_1$ is the ray that contains $r$ and $V_2 ... V_k$ are the other rays. A split with border $B$ can not overlap any $V_i$. As $B$ touches $V_2 ... V_k$ only one split has border $B$, namely $(V_0 \cup V_1, V_2 \cup ... V_k)$. It is weak since it is crossed by split $(V_2 \cup V_0, V_1 \cup V_3 \cup ... V_k)$.

Conversely let $B$ be a strong border (i.e. in $\mathcal{B}_h \cap \mathcal{B}_h^{\perp}$) that is border of no strong split. We shall prove that $B$ is a bad star border. As each border is the border of at least one split, then $B$ must be the border of at least two splits, all weak.

From Proposition 5 we have that these weak splits are either in Clique or Star configuration into $V_0 ... V_k$. First, let us study the case where the split is either a Clique configuration or a Good Star configuration and without loss of generality we can assume that $r \in V_0$ (in the good star case it means that $V_0$ is the center of the star). For $i > 0$, let $V_i'$ be the vertices of $V_i$ incident with $V_0$. Then any union of $V_i'$ is a border. Therefore only the maximal union $B = V_1' \cup ... V_k'$ is a strong border. Then $B \cap V_0 = \emptyset$ but for $i > 0$ $B \cap V_i \neq \emptyset$. Split $(V_0, V_1 \cup ... V_k)$ has border $B$ and is strong, a contradiction.

So we are left with the Bad Star configuration case (we still assume $V_0$ to be the center, so $r \notin V_0$). Let $B$ be a bad star border contained in a larger bad star border $B'$. $B'$ is not connected in $G[\geq h]$ and each connected component corresponds to a ray of the star. Some of them belong to $B$, other not. A combination of such rays overlaps $B$, which can not be a strong border. So $B$ is maximal with respect to inclusion. The last claim of the proposition comes from fact that a bad star border has no neighbour inside $G[h]$ (these neighbours would have to be in the center of the star and therefore be connected to some elements of the upper ray $V_1$, which is not possible since these elements belong to $G[h-2]$). ■

# 4 Split Decomposition - Algorithm

In this section, we show how the theory developed in Section 2 combined with the algorithm for computing the orthogonal of a family (Theorem 2) allow us to design a $O(|E|)$-algorithm to produce Cunningham's tree decomposition. The exact complexity analysis is postponed to Section 5.

## 4.1 Building the tree decomposition - General Approach

Our algorithm constructs Cunningham's tree of strong splits in a step by step "bottom up" approach. At each step of our algorithm we produce a forest $\mathcal{F}_h$ of rooted trees which roughly represents all labelled split bottoms at distance at least $h$, for $h$ going down to 1.

From now on, we call *internal nodes* of this forest the nodes which are neither leaves, nor roots. As in Cunningham's tree, the leaves of our forest are labelled with vertices of the graph (each vertex is associated to at most one leaf), and each non-leaf node is associated to a subset of $V(G)$ which are the labels of the descendants of this node. We say that this node *represents* this set. Sometimes, to simplify notations and if no confusion may occur, we will identify a node with the set it represents.
The following invariants are maintained after processing layer $h$:

**Invariant 1** The leaves of $\mathcal{F}_h$ are exactly labelled by the vertices at distance greater or equal to $h$ (vertices of $G[\geq h]$).

**Invariant 2** Each strong split bottom at distance greater or equal to $h$ is represented by one node of $\mathcal{F}_h$.

**Invariant 3** Each internal node of $\mathcal{F}_h$ represents a strong split bottom and is given with its correct label Star, Prime or Clique.

**Invariant 4** Each root of $\mathcal{F}_h$ represents either a connected component of $G[\geq h]$ that is not a split bottom, or a connected split bottom (it that case it is labelled either Clique or Prime) or a split bottom that is the union of several such connected components (and in that case it is labelled Star).

The algorithm constructs the forest $\mathcal{F}_h$ from $\mathcal{F}_{h+1}$ by adding new leaves (vertices of $G[h]$) and new nodes. For $h = n$ the initial forest is empty. The process ends for $h = 0$ when $r$ is added.

Notice that $(V(G) \setminus r)$ is a strong split bottom and therefore has to be represented by a node $P$ in $\mathcal{F}_1$ (Invariant 2). This implies that the forest $\mathcal{F}_1$ is in fact a unique tree and the node $P$ is the root of this tree. By adding $r$ as a leaf attached to $P$, $P$ and all internal nodes represents a strong split bottom (Invariant 3), and all split bottoms are represented by a node (Invariant 2). Therefore this last tree is that of Cunningham.

*Thus, we only need to show how to construct $\mathcal{F}_h$ from $\mathcal{F}_{h+1}$ while preserving these four invariants.*

The following point is important. Assume that we want to process layer $h$ and compute $\mathcal{F}_h$ from $\mathcal{F}_{h+1}$. Since we maintain Invariant 2, *i.e.* each split bottom at distance $h$ is represented by a node of the forest, and since split bottoms at distance $h+1$ or more are already represented, *we only need to care about split bottoms at distance exactly $h$*, *i.e.* split bottoms with borders included in the layer $G[h]$. Thus the leaves we add are exactly the vertices of $G[h]$, and, as explained below, the internal nodes we add correspond exactly to the bottoms.

## 4.2 Recursive Computation of $\mathcal{F}_h$

We explain in this section how to build the forest $\mathcal{F}_h$ from the two forests $\mathcal{F}_{h+1}$ and the forest of borders $\mathcal{B}_h$.

The first step is to slightly transform $\mathcal{B}_h$ to consider the connectivity inside layer $h$ (maintaining Invariant 4). Let us call *h-component* the intersection of a connected component of $G[\geq h]$ with $G[h]$. We build the forest $\mathcal{B}'_h$ the following way: each tree of $\mathcal{B}_h$ is incorporated in $\mathcal{B}'_h$. Then, for each h-component that contains the roots of at least two trees, we create a new node corresponding to this h-component with these roots as children. Furthermore we cast the type of all *complete* node of $\mathcal{B}_h$. Let $N$ be such a complete node and $S_1$, $S_2$ its two first sons, and pick two vertices $x_1 \in S_1$ and $x_2 \in S_2$. If they are adjacent, then $N$ is relabelled *clique*, otherwise it is relabelled *star*. If $N$ is *star* and has a parent then the center of the star is that parent (else, it shall be defined later). For correctness of the typing see the proof of Invariant 3.
We need the following result to properly state the algorithm.

**Proposition 8** *Let $C$ be a connected component of $G[> h]$. Then $N_h(C)$ is contained in at least one node of $\mathcal{B}'_h$ (i.e. in the vertex-set of $G[h]$ represented by this node).*

**Proof** All elements of $N_h(C)$ belong to the same $h$-component. ■

**Notations:** Notice that each forest is defined by a *parent* relation between nodes, undefined for roots. We perform three kinds of operations for creating $\mathcal{F}_h$:

- *Merging* Node $A$ with Node $B$ means setting each child of $A$ as a new child of $B$ and removing $A$ (notice that it is not commutative)
- *Linking* $A$ to $B$ sets $parent(A) := B$.
- *Adding a parent* to node $A$ consists in creating a new node $B$, setting $parent(B) := parent(A)$ and then $parent(A) := B$.

**Remark:** Thanks to Invariant 4, we know that a root of $\mathcal{F}_{h+1}$ represents either one connected component of $G[> h]$ or a union of connected components that have the exact same neighborhood in $G[h]$. Therefore, all the results of Sections 4.2 and 4.3 are perfectly valid if we replace the family of connected components of $G[> h]$ by the family of roots of $\mathcal{F}_{h+1}$. Then we use the terminology of Proposition 6 for a root $R$ of $\mathcal{F}_{h+1}$ (instead of a component $C$) and call it Type 1, Type 2, or Type 3 with respect to some given border in $\mathcal{B}_h$.

**Algorithm:**

1. Compute $\mathcal{B}_h$
2. Compute $\mathcal{B}'_h$
3. For each root $R$ of $\mathcal{F}_{h+1}$, let $B$ be the lowest node of $\mathcal{B}'_h$ such that $N_h(R) \subset B$.
   (a) If $R$ is a not a split bottom ($R$ is just a connected component of $G[\geq h]$), then **merge** $R$ with $B$.
   (b) Else if $R$ is of Type 2 with respect to $B$, or if $B$ is not a split border, then **link** $R$ to $B$. If $R$ was labelled Star, then orient the edge from $R$ to $B$.
   (c) Else if $R$ is of Type 3 with respect to $B$ and $R$ is labelled Star then **add a parent** $P$ to $B$ and **merge** $R$ with $P$. Label $P$ star and orient the edge $PB$ from $P$ to $B$.
   (d) Else ($R$ is of Type 3 with respect to $B$ but not labelled Star) **add a parent** $P$ to $B$ and **link** $R$ with $P$. Label $P$ star and orient the edge $PB$ from $P$ to $B$.

## 4.3 Correctness

As noted before, we just need to prove that Invariants 2,3 and 4 are still true after the update.

**Invariant 2:**
The only nodes of $\mathcal{F}_{h+1}$ that are destroyed during the update are the roots that we merge. We do this only in two cases. Either when the root is either a connected component but not a split bottom, and in that case it is not problem (and it is needed by Invariant 3). Or when the root is labelled star, which by Invariant 4 implies that it represents a disconnected split bottom, and its neighborhood in $G[h]$ is a split border. This is exactly the case of a bad star, and thanks to Proposition 7, we knew this node had to be deleted (they don't represent strong bottoms). So we preserved all strong split bottom of distance at least $h+1$.

Now we need to prove that this invariant is true also for strong split bottoms at distance $h$. Such a strong split bottom $S$ has its border $B_S$ in $\mathcal{B}_h \cap \mathcal{B}_h^{\perp}$, which means that it is represented by a node $N$ in $\mathcal{B}_h$. Thanks to Corollary 2, we know that at most two strong split bottoms with border $B_S$ can exist, $B_S$ along with components of Type 2, and $B_S$ along with components of Type 2 and 3. With respect to $B_S$ every root of $\mathcal{F}_{h+1}$ is either of Type 1, 2, or 3. Type 1 components are never placed by the algorithm under node $N$, since their neighborhood is not included in $B_S$. Type 2 components are always put under $N$, since $N$ is either $B$ or an ascendant of $B$, where $B$ is the smallest node containing their neighborhood. Therefore, in $\mathcal{F}_h$ the vertices below node $N$ are exactly $B_S$ along with Type 2 components. Finally if any Type 3 components exist for $B_S$, then the algorithm creates a new node $P$ in $\mathcal{F}_h$. The vertices below this node $P$ in the end are exactly $B_S$ with all components of Type 2 and of Type 3.

**Invariant 3:**
Let $N$ be an internal node of $\mathcal{F}_h$. There are three possible cases coming from the update algorithm:

1. $N$ comes from a node of $\mathcal{F}_{h+1}$ (either internal or a root that has not been merged).
2. $N$ is created by **add a parent**.
3. $N$ comes from a node of $\mathcal{B}'_h$.

If $N$ comes from an internal node of $\mathcal{F}_{h+1}$, since the subtree rooted in $N$ has not been modified by the update and by Invariant 3 at the previous step, we know that it represents some strong split bottom at distance greater than $h$. If $N$ comes from a root of $\mathcal{F}_{h+1}$ and has not been merged (cases $(b)$ and $(d)$ of the algorithm), then it means that it represents a maximal split bottom of distance greater than $h$ that is not a bad star bottom, since case $(a)$ deals with $R$ not split bottom and case $(c)$ with $R$ bad star border with $\geq 3$ rays. Therefore by Proposition 7 it represents a stong split bottom.

If $N$ is a node that was created by an **add a parent** operation, then it means that it represents a split border along with all components of Type 2 and 3 with respect to it. This is also the case of a strong split bottom.

Eventually, if $N$ comes from a node of $\mathcal{B}'_h$ and is internal, then $N$ represents a border along with its Type 2 Components. Indeed, the only nodes in $\mathcal{B}'_h$ that do not represent borders are the roots added to $\mathcal{B}_h$ during the creation of $\mathcal{B}'_h$ but the algorithm never adds a parent to those, so they remain roots. Therefore we know that $N$ represent a split bottom, and we need to prove that it represents a strong one. Thanks to Proposition 7, we know that if a split bottom is not strong but has a border in $\mathcal{B}_h \cap \mathcal{B}_h^\perp$, then this border must be a root of $\mathcal{B}'_h$. Notice also that the update algorithm never adds a parent to such a root since it only does if there exists a component of $G[> h]$ of type 3 with respect to it (and therefore connecting all of them), but this is not possible in the case of a bad star border since every ray is in a different connected component of $G[\geq h]$. Therefore, these split bottoms can only be represented by roots of $\mathcal{F}_h$.

So each internal node represents a strong split bottom. We now just have to prove that the node labelling is correct. From Proposition 5 we know the labels have to be Prime, Clique or Star. If an internal node comes from $\mathcal{B}_h$, then it had a Prime or Complete label. A split bottom is Prime if and only if no union of its children defines a split bottom, and therefore if and only if its border is a prime border. Moreover, recall that for a given split bottom, if some union of its children defines a split bottom (it is labelled Complete in $\mathcal{B}_h$), then any union does and in that case it is Clique if and only if it is connected, and Star if every children defines a connected component. These conditions are exactly the one we apply when transforming the labels of $\mathcal{B}_h$ into the ones of $\mathcal{B}'_h$.
If the internal node comes from $\mathcal{F}_{h+1}$, since labels are unmodified, the validity is guaranteed by the Invariants at the previous step.
Eventually, if the internal node comes from an **add a parent** operation, then it means that it represents a star split bottom (it has a component of Type 3) whose center is the part containing the border and is thus labelled accordingly.

**Invariant 4:**
By construction, roots of $\mathcal{F}_h$ are of three kind:

1. either they come from a root of $\mathcal{B}_h$,
2. or they were added during the construction of $\mathcal{B}'_h$ (because of a $h$-component),
3. or they were added on some root of the first kind during the update because of Type 3 components (we do not apply **add a parent** to nodes that represent $h$ components).

In the first or third case, it represents a split bottom and is well-labelled for the same exact reasons as for internal nodes (see proof of Invariant 3). In the second case, it is not a split bottom and thus receives no label. Note that from the definition of $\mathcal{B}'_h$, if we denote by $A$ a root of $\mathcal{B}'_h$ and $C$ is a connected component of $G[> h]$, either $C$ is of Type 2 or 3 with respect to $A$, or $N_h(C) \cap A = \emptyset$. It clearly implies that when $A$ becomes a root of $\mathcal{F}_h$ it represents a connected component of $G[\geq h]$.

# 5 Split Decomposition Algorithm - Complexity Analysis

Let $E_h$ denote the set of edges in $G[h]$ and $E_{h,h+1}$ the set of edges between $G[h]$ and $G[h+1]$. The efficiency of the whole algorithm relies on the fact that the update algorithm on layer $h$ runs in time proportional to $|H| + |E_{h-1,h}| + |E_h| + |E_{h,h+1}|$, which is proved below.

Indeed, computing $\mathcal{B}_h$ and $\mathcal{B}'_h$ efficiently relies on the linear time computation of orthogonals, modular decomposition and connected components. In the general case, the former can be handled using the algorithm of McConnell (see Section 2). It is not difficult to implement this algorithm in such a way that, in addition to the tree representing the orthogonal of a given family of subsets, one can get for each element of the family a pointer to the smallest element of the orthogonal that contains it, which is what we need during the recursive computation of $\mathcal{F}_h$ . However, the sizes of the families of Section 4 might not be linear in the size of the underlying graph. We show below how to compute families with the same orthogonals. The modular decomposition aspects are explained in Section 5.3. Computing the connected components of $G[\geq h]$ for all $h$ is also clearly linear in the number of edges of the graph.

Computing $\mathcal{F}_h$ from $\mathcal{F}_{h+1}$ and $\mathcal{B}'_h$ can be done in linear time using classical algorithmic operations on trees, since for all roots $R$ the lowest node $B$ in $\mathcal{B}'_h$ can be identified in the number of leaves belonging to $G[h+1]$ of the subtree rooted by $R$ using a bottom up marking approach.

### 5.1 Computing $\mathcal{M}$.

**Theorem 6** *[26, 17, 5] Modules of $G[< h]$ form a partitive family and therefore can be represented as a tree (Theorem 1) where each node represents a strong module. Computing this labelled tree is called* modular decomposition *and requires a time proportional to $|E_{h-1,h}|+|E_h|$.*

We compute a representation of $\mathcal{M}$, *i.e.* the modules of $G[\leq h]$ belonging to $G[h]$, which is a partitive family on $G[h]$ up to the fact that $H$ itself might not belong to the family therefore leading to a modular decomposition forest. Notice that these modules are defined only by their adjacencies with layers $h$ and $h-1$. We first compute the modular tree decomposition of the modules of $G[h, h-1]$ and marking each leaf belonging to $G[h]$ that is added to the representation. We then perform a bottom-up selection, each node *Prime* or *Complete* of which all children are marked is also marked and an exact copy of this node is added to the tree representation of $\mathcal{M}$. If all children of a node are not marked, if the node if *Prime* the bottom-up process stops from this node, otherwise the *Complete* node is added with the marked children only and the bottom-up process continues from it. We eventually replace each *Complete* nodes with a single child by this child.
An important property for the time complexity of our algorithm is:

**Proposition 9** $\|\mathcal{M}\| = O(|E_{h-1,h}| + |E_h| + |H|)$.

**Proof** This fact directly derives from that the total sum of all elements of all strong modules of a graph $G$ is $O(n+m)$ as proved in [16]. ■

### 5.2 Efficient Computation of Overlaps in Two Particular Cases

We describe here two tools that allow us to efficiently compute orthogonals in two particular cases. Both are of use for assessing the linear complexity of the algorithm.

**Lemma 2** *Let $V = \{x_1, x_2, \ldots, x_p\}$ be a finite set. Let $\mathcal{A} = \{\{x_i, x_{i+1}\},\ i = 1 \ldots p,$ and $x_{p+1} = x_1\}$. Then $\mathcal{A}^\perp = (2^V)^\perp = \{\{x_i\},\ i = 1 \ldots p\}$*

As a consequence, consider a prime node of a partitive tree with children $A_1, \ldots A_p$. We know that it is the orthogonal of its associated *Complete* node, *i.e.* the family of all possible unions of sets $A_i$. What the lemma says is that it is also the orthogonal of the family with $p$ elements of this type: $A_i \cup A_{i+1}$, and this family is much smaller than all possible unions, since it has norm twice the one of the $A_i$. We call $\{A_1 \cup A_2, A_2 \cup A_3 \ldots A_p \cup A_1\}$ the *circulant* family associated to the $A_i$. Schematically, this can be graphically represented as the "equality" Figure 6.

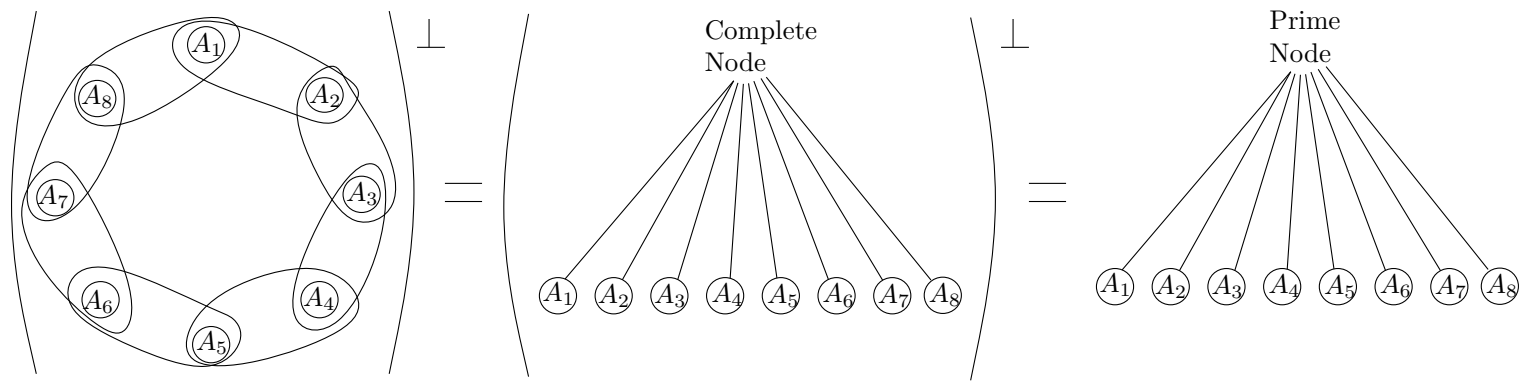

*Figure 6: A circulant family and its orthogonal.*

**Proposition 10** *Let $\mathcal{F}$ be a family of subsets of $V$. Given the partitive tree $T(\mathcal{F}^{\perp})$ representing $\mathcal{F}^{\perp}$, it is possible to construct a family $\mathcal{H}$, such that $\mathcal{F}^{\perp} = \mathcal{H}^{\perp}$, and such that the size of $\mathcal{H}$ and time needed to do this calculation are proportional to the size of this tree, that is $\|\mathcal{F}^{\perp\perp} \cap \mathcal{F}^{\perp}\|$.*

**Proof** The family $\mathcal{H}$ is simply the family containing the sets represented by the nodes of the tree (i.e. elements of $\mathcal{F} \cap \mathcal{F}^{\perp}$) plus, for each prime node of this tree, the circulant family associated to the children of this node. ■

We use this result to prove another proposition used in the next section.

**Proposition 11** *Let $V$ be a finite set, and $X$ be a subset of $V$. Assume that $\mathcal{F}$ is a family of subsets of $X$ such that $X \in \mathcal{F}$. Let us define a new family $\mathcal{H}$ by $\mathcal{H} = \mathcal{F} \cup \{X \setminus F,\ F \in \mathcal{F}\}$. Then in time $O(\|\mathcal{F}\|)$, it is possible to compute a family $\mathcal{H}'$ such that $\mathcal{H}'^{\perp} = \mathcal{H}^{\perp}$ and which size is in $O(\|\mathcal{F}\|)$.*

**Proof** Let $P_1, \cdots, P_t$ be the equivalence classes of the following relation on elements of $V$: $x$ and $y$ are equivalent if they belong tho the exact same members of the family $\mathcal{H}$. The sets $P_i$ thus form a partition of $V$ and we define $\mathcal{P}$ to be the family composed of all $P_i$ and all possible unions of $P_i$. We prove below that $\mathcal{P}^{\perp} = \mathcal{H}^{\perp}$.

Let $A \in \mathcal{H}^{\perp}$, then $A$ is either included in a $P_i$ or $A$ is a subset of to $V \setminus X$. Therefore $A \in \mathcal{P}^{\perp}$. Conversely, assume now that $A \in \mathcal{P}^{\perp}$. As all unions of $P_i$ belong to $\mathcal{P}$, $A$ is either included in a $P_i$ or a subset of $V \setminus X$. Thus $A \in \mathcal{H}^{\perp}$.

The size of the family $\mathcal{P}$ might however be too large for our purpose. Thus, instead of considering all unions of all $P_i$, we use Proposition 10 and consider instead the family $\mathcal{H}'$ of circulant unions $P_i \cup P_{i+1}$. We therefore insure that $\mathcal{H}'^{\perp} = \mathcal{H}^{\perp}$ and that $\|\mathcal{H}'\| = \|\mathcal{F}\|$

The time complexity of the construction of $\mathcal{H}'$ relies on the efficiently of building $P_1, \cdots, P_t$. We use partition refinement that can be done by the very simple following process: let $\mathcal{U}$ be a family on $V'$, containing only $X$ at the beginning. We consider successively each set $Y \neq X$ in $\mathcal{F}$ as pivot. Let $C \in \mathcal{U}$ such that $C = C' \cup C''$, with $C' \not\subseteq Y$, $C'' \subseteq Y$, $C' \neq \emptyset$, and $C'' \neq \emptyset$. We only replace $C$ by the two sets $C'$ and $C$" in $\mathcal{U}$. At the end of this process, $\mathcal{U}$ is the partition of $P_i$ of $X$ we aim for. This refinement procedure can be implemented in $O(\|\mathcal{F}\|)$ using a structure based on an augmented array [2] or based on an ad-hoc double linked list [12]. ■

### 5.3 Computing Efficiently The Family of Borders

In this section, we show how to use the tools of sections 2 and 3 to compute the family of borders $\mathcal{B}_h$ efficiently, that is with a linear (with respect to the number of edges of the graph) time complexity. Theorem 5 asserts that

$$\mathcal{B}_h \cup \{H\} = (\mathcal{M}^{\perp} \cup \mathcal{V})^{\perp}.$$

Of course, we want to apply the orthogonal algorithm of McConnell ([15], see Section 2). But if we do this directly on the family $(\mathcal{M}^{\perp} \cup \mathcal{V})$ the time complexity will be higher than what we want, because this family is a bit too large (for instance because of the complements of neighborhoods in family $\mathcal{V}$). To avoid this issue, we are using the reduction tools of Section 5.2.

Using the notations of Section 3.2,

$$B_h \cup \{H\} = (\mathcal{M}^{\perp} \cup \mathcal{V})^{\perp} = (\mathcal{M}^{\perp})^{\perp} \cap \mathcal{V}_1^{\perp} \cap \mathcal{V}_2^{\perp} \cap \ldots \cap \mathcal{V}_k^{\perp} \tag{1}$$

The forest representing the family $\mathcal{M}$ can be calculated in time $O(|E_h| + |E_{h-1,h}|)$. Then the tree representing $\mathcal{M}^{\perp}$ can be obtained by simply swapping *Prime* and *Complete* nodes as stated in Proposition 3. Now, as the total size of the family is linear (Proposition 9), using Proposition 11 it is possible to construct in time $O(|H| + |E_h| + |E_{h-1,h}|)$ a family $\mathcal{N}$ such that

$$\mathcal{N}^{\perp} = (\mathcal{M}^{\perp})^{\perp} = \mathcal{M}$$

Furthermore, using Proposition 11, it is possible, in time proportional to the sum of the sizes of the sets in $\{N(C_i) \cap H\} \cup \{N(x) \cap H, x \in C_i\}$, to compute a family $\mathcal{W}_i$ such that:

$$\mathcal{W}_i^{\perp} = \mathcal{V}_i^{\perp}.$$

Let us define $\mathcal{W}$ to be the union of all $\mathcal{W}_i$. It is important to note that $\|\mathcal{N}\|$ is $O(|H| + |E_h| + |E_{h-1,h}|)$ while $\|\mathcal{W}\|$ is $O(|H| + |E_h| + |E_{h,h+1}|)$. Since

$$\begin{aligned} B_h \cup \{H\} &= (\mathcal{M}^{\perp})^{\perp} \cap \mathcal{V}_1^{\perp} \cap \mathcal{V}_2^{\perp} \cap \ldots \cap \mathcal{V}_k^{\perp} \\ &= \mathcal{N}^{\perp} \cap \mathcal{W}_1^{\perp} \cap \mathcal{W}_2^{\perp} \cap \ldots \cap \mathcal{W}_k^{\perp} \\ &= (\mathcal{N} \cup \mathcal{W})^{\perp}, \end{aligned}$$

Thus, we are able to compute a tree representation of $B_h \cup \{H\}$ by computing in extension $\mathcal{N} \cup \mathcal{W}$ in a total time $O(|H| + |E_h| + |E_{h-1,h}| + |E_{h,h+1}|)$ and by using Theorem 2 in the same time. We just proved the following theorem:

**Theorem 7** *The partitive tree representing split borders at distance $h$ can be calculated in $O(|V(G[h])| + |E_h| + |E_{h-1,h}| + |E_{h,h+1}|)$ time.*

Doing this for all $h$, we get an algorithm that is linear with respect to the total number of edges in the graph.

### Acknowledgment

We wish to thank Vincent Limouzy for his participation to the earlier studies we did on the subject.